      [1999/12/01 v1.4c Il Nuovo Cimento]
\documentclass{cimento}
\usepackage{epsfig,psfrag,graphicx,bm,color,verbatim}
\usepackage{amsmath}
\usepackage{amsfonts}
\usepackage{amssymb}

\begin{document}
\title{Measurement of anisotropies in the large-­scale diffuse gamma-­ray emission}
\author{G.~G\'omez-Vargas,
on behalf of the Fermi-LAT collaboration, and E.~Komatsu \\
}
\instlist{\inst{} Departamento de F\'{\i}sica Te\'{o}rica, and Instituto de F\'{\i}sica Te\'{o}rica UAM/CSIC,\\ Universidad Aut\'{o}noma de Madrid, Cantoblanco, E-28049, Madrid, Spain \\
Istituto Nazionale di Fisica Nucleare, Sez. Roma Tor Vergata, Roma, Italy}

\maketitle

\begin{abstract}
We have performed the first measurement of the angular power
spectrum in the large­-scale diffuse emission at energies from 1-­50
GeV. We compared results from data and a simulated model in order to identify significant differences in anisotropy properties. We found angular power above the photon noise level in the data at multipoles greater than $\sim 100$ for energies $1 \lesssim E \lesssim 10$ GeV.  The excess power in the data suggests a contribution from a point source population not present in the model.

\end{abstract}

\section{Introduction}
The Fermi Gamma-Ray Telescope, launched on June 11th 2008 from Cape Canaveral, performs gamma-ray measurements over the whole celestial sphere. Its main scientific instrument the Large Area Telescope (LAT) measures the tracks of the electron and positron that result when an incident gamma-ray undergoes pair conversion in a thin, tungsten foil, and measures the energy of the subsequent electromagnetic shower that develops in the telescope's calorimeter. Some \textit{Fermi}-LAT specifications are: Energy range from 20MeV to $\sim$300GeV, angular resolution $\sim$0.1 deg above 10 GeV, field of view (FoV) $\sim$2.4sr, and uniform sky exposure of $\sim$30 minutes every 3 hours. Detailed descriptions of the \textit{Fermi}-LAT telescope can be found in \cite{Atwood:2009ez}.

One of the key science targets of the Fermi mission is diffuse gamma-ray emission. Its main component is correlated with Milky Way structures, the galactic emission, arising from interactions of high-energy cosmic rays with the interstellar medium and the interstellar radiation field. A fainter component considered to have an isotropic or nearly isotropic distribution on the sky, the so-called extragalactic emission, has been observed. This observation is based on the modelization of galactic component, detected \textit{Fermi}-LAT sources and the solar gamma-ray emission \cite{Abdo:2010nz}. Also there is a contribution from populations of sources, of various kinds, including blazars, pulsars, SNR, and possibly dark matter (DM) structures, not yet detected due to \textit{Fermi}-LAT spatial resolution and photon statistics. The angular distribution of photons in the diffuse gamma-­ray background contains information about the presence and nature of these unresolved source populations (USP). Fluctuations on small scales may originate from USP if they are different from those expected from the Poisson noise due to finite statistics.

Recent studies have predicted the contributions to the angular power spectrum (APS) from extragalactic and galactic DM annihilation or decay, e.g. \cite{Ando:2005xg, Ando:2006cr, SiegalGaskins:2008ge, SiegalGaskins:2009ux, Ando:2009fp,Fornasa:2009qh,Zavala:2010pw,Taoso:2008qz}. A detailed \textit{Fermi}-LAT sensitivity study of anisotropies from DM annihilation has been presented in \cite{Cuoco:2010jb}.

I present the results of an anisotropy analysis of the diffuse emission measured by the \textit{Fermi}-LAT. We calculate the angular power spectrum of the emission from $\sim 22$ months of Fermi data and of the emission from a simulated model (galactic diffuse emission, 11-month sources from Fermi catalog and isotropic emission), and compare the results from the data and model in order to identify significant differences in anisotropy properties.

\section{The angular power spectrum (APS) as a metric for anisotropy}

We consider the APS $C_l$ of intensity fluctuations,

\begin{equation}\label{delta}
\delta I(\psi)=\frac{I(\psi)-\langle I\rangle}{\langle I\rangle},
\end{equation}

where $I(\psi)$ is the intensity in the direction $\psi$. The APS is given by $C_l=\langle \vert a_{lm}\vert^ 2\rangle$, where $a_{lm}$ are determined by expanding (\ref{delta}) in spherical harmonics, $\delta I(\psi)=\sum_{l,m}{a_{lm}Y_{lm}}.$

The $1-\sigma$ statistical uncertainty in the measured APS is given by
\begin{equation}
\delta C_l = \sqrt{\frac{2}{(2l+1)\Delta l f_{sky}}}\left(C_l+\frac{C_N}{W_l^ 2} \right)
\end{equation}
where $W_l=$exp$(-l^2\sigma^2_b/2)$ is the window function of a Gaussian beam of width $\sigma_b$. $f_{sky}$ is the fraction of the sky observed and $\Delta l$ multipole bins. The noise power spectrum $C_N$ is the Poisson noise, $C_N=(4\pi f_{sky})/N_{\gamma}$, where $N_{\gamma}$ is the number of photons observed.

 Predicted values of $C_l$ at $l=100$ of various USP cover a large range, e.g., $\sim 1\times 10^{-4}$ for blazars \cite{Ando:2006mt}, $\sim 1\times 10^{-7}$ for starforming galaxies \cite{Ando:2009nk}
, and $\sim 1\times 10^{-4}$ to $1$ for DM \cite{Ando:2005xg, Ando:2006cr, SiegalGaskins:2008ge, SiegalGaskins:2009ux,Ando:2009fp,Fornasa:2009qh,Zavala:2010pw,Taoso:2008qz}.
\section{Method}
\begin{itemize}
\item Select regions of the sky which are relatively clean
	\begin{list}{-}{}
	\item mask sources in the 11-month catalog within a 2 deg radius
	\item mask the galactic plane $\vert b\vert<30$deg
	\end{list}
\item    Calculate angular power spectrum of the data in several energy bins using
   the HEALPix package \cite{Gorski:2004by}.
­\item Focus on multipoles greater than 100 (angular scales $\lesssim 1-2^{\circ}$), because
   the contamination from Galactic diffuse is likely to be small.
­\item Compare results from data and simulated model to identify significant
   differences in anisotropy properties.
­\item Error bars on points indicate $1-\sigma$ statistical uncertainty in the measurement; systematic uncertainties are NOT included.

\end{itemize}

Data from $\sim 22$ months of diffuse class events in the energy range 1GeV to 50GeV were analyzed. We used P6\_\,V3 instrument response, for data and simulations. Maps have been binned into order 9 HEALPix.

The  simulated data are produced using the \texttt{gtobssim} routine, part of the Fermi Science Tools package. We used current background models released by the Fermi collaboration\footnote{http://fermi.gsfc.nasa.gov/ssc/data/access/lat/BackgroundModels.html} and 1-year point source catalog\footnote{http://fermi.gsfc.nasa.gov/ssc/data/access/lat/1yr\_\,catalog/}.

\section{Results and conclusions }
Plots of Fig. 1. show the APS of the data and the default model (Galactic diffuse model + 11 month source catalog + isotropic) in different energy ranges. These figures show at what energy ranges and multipole ranges the APS of the data and the model differ, as well as where each of these is consistent with the photon noise level.

We have found that at multipoles greater than $\sim 100$ the excess power in the data suggest a contribution from a point source population not present in the model. Also, at large angular scales ($l<100$) angular power above the noise is seen in the data and model, probably due to contamination from the galactic diffuse.

Due to decreasing photon statistics, the amplitude of anisotropies detectable by this analysis decreases with increasing energy. For this reason, the non-detection of power above the noise level at $10$-$­50$ GeV does not exclude the presence of anisotropies at the level of those detected at $1$-­$10$ GeV.
\begin{center}
\begin{figure}[htb]\label{plots}
\includegraphics[width=0.45\textwidth]{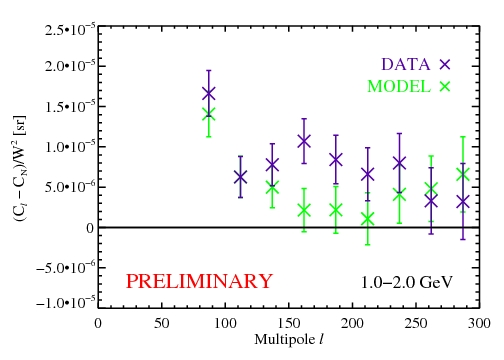}\quad
\includegraphics[width=0.45\textwidth]{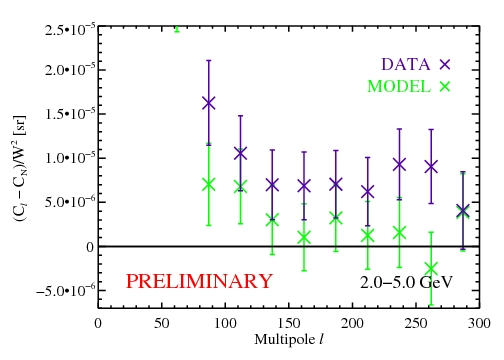}\quad
\includegraphics[width=0.45\textwidth]{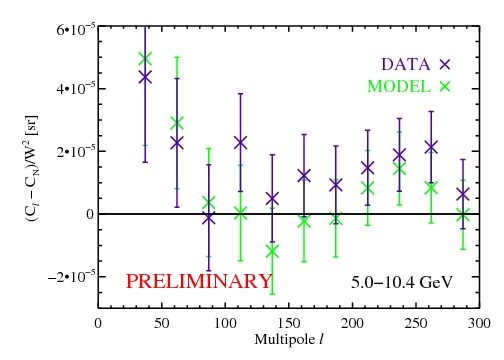}\quad
\includegraphics[width=0.45\textwidth]{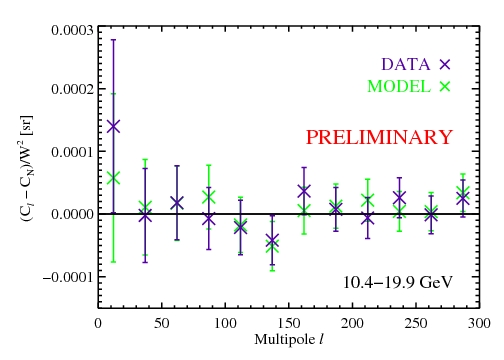}\quad
\includegraphics[width=0.45\textwidth]{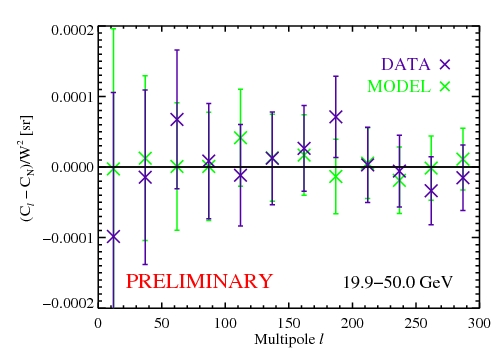}
\caption{Fluctuation APS of the data and the default model (Galactic diffuse model + 11 month source catalog + isotropic) in different energy ranges. }
\end{figure}
\end{center}

\acknowledgments
The \textit{Fermi}-LAT Collaboration acknowledges support from a number of agencies and institutes for both development and the operation of the \textit{Fermi}-LAT as well as scientific data analysis. These include NASA and DOE in the United States, CEA/Irfu and IN2P3/CNRS in France, ASI and INFN in Italy, MEXT, KEK, and JAXA in Japan, and the K.~A.~Wallenberg Foundation, the Swedish Research Council and the National Space Board in Sweden. Additional support from INAF in Italy and CNES in France for science analysis during the operations phase is also gratefully acknowledged. G.G.V thanks the support of the Spanish MICINN’s Consolider-Ingenio 2010 Programme under grant MultiDark CSD2009-00064, MICINN under grant FPA2009-08958, the Community of Madrid under grant HEPHACOS S2009/ESP-1473, and the European Union under the Marie Curie-ITN program PITN-GA-2009-237920.

\end{document}